# Classification of meaningful and meaningless visual objects: a graph similarity approach

Ahmad Mheich[1], Mahmoud Hassan[2,3], Fabrice Wendling[2,3]

[1] Télécom SudParis, EPH, Évry, F-91000, France
[2] INSERM, U1099, Rennes, F-35000, France
[3] Université de Rennes 1, LTSI, F-35000, France

*Abstract*— Cognition involves dynamic reconfiguration of functional brain networks at sub-second time scale. A precise tracking of these reconfigurations to categorize visual objects remains elusive. Here, we use dense electroencephalography (EEG) data recorded during naming meaningful (tools, animals…) and scrambled objects from 20 healthy subjects. We combine technique for identifying functional brain networks and recently developed algorithm for estimating networks similarity to discriminate between the two categories. First, we showed that dynamic networks of both categories can be segmented into several brain network states (times windows with consistent brain networks) reflecting sequential information processing from object representation to reaction time. Second, using a network similarity algorithm, results showed high intra-category and very low inter-category values. An average accuracy of 76% was obtained at different brain network states.

*Keywords—Dense-EEG source connectivity, Brain networks, network simialrity*

## I. INTRODUCTION

The human brain is a complex network of functionally interconnected distant brain regions. There is increasing evidence that the dynamic reconfiguration of this network is a crucial aspect to understand information processing in the human brain [1]. Recently, many studies showed that the dynamic reconfiguration of the brain networks is related to different brain functions such as learning [2] and disease progression [3] at different time scales mainly minutes, hours and days [3]. A precise tracking of these reconfigurations at sub-second time scale is challenging. One of these fast time scale cognitive activities is the object recognition and categorization. This cognitive function occurs in sub-second time period.

The existed literature related to object categorization is mainly focused on the time/frequency analysis of the evoked responses. It was used to differentiate between non-natural vs. natural [4], [5] made vs. non-made [4] and human vs. non-human [6] objects. To what extent the semantic category of the visual stimuli can be related to the dynamic reconfiguration of the functional brain networks is still unclear.

To tackle this issue, we collected dense-electroencephalography (EEG, 256 channels) data from subjects performing a cognitive task corresponds to name two categories of pictures (meaningless and meaningful). Functional brain networks for each object in both categories were estimated using 'dense-EEG source connectivity' method [7], [8]. These brain networks were then segmented into brain network states (BNS) using a $k$-means clustering method [9]. At each BNS, the similarity between networks were computed using an algorithm called 'SimNet' [10], [11]. The main feature of SimNet is the taking into consideration the spatial location of nodes in order to find the similarity scores between graphs, a key feature in the brain network analysis context. The results showed a high performance of the proposed network similarity-based approach to differentiate between visual object categories.

## II. MATERIALS AND METHODS

### A. Experimental protocol

Twenty healthy volunteers (10 women and 10 men; mean age, 23 y) with no neurological disease were involved in this study. 79 meaningful and 40 scrambled pictures were displayed and the participants were asked to name them. All pictures were shown as black drawings on a white background. Errors in naming were removed for the further analysis and the signals of one participant were eliminated as data were very noisy. The brain activity was recorded using dense-EEG, 256 electrodes, system (EGI, Electrical Geodesic Inc.) with a 1 kHz sampling frequency providing a high temporal resolution and band-pass filtered between 3 and 45Hz. Each trial was visually inspected, and epochs contaminated by eye blinking, movements or any other artifacts were rejected and excluded from the analysis performed using the EEGLAB and Brainstorm open source toolboxes [12], [13].

### B. Functional connectivity analysis

Functional connectivity matrices were estimated using EEG source connectivity method [8]. Briefly, the method includes two main steps: i) solving the EEG inverse problem to reconstruct cortical sources and ii) measuring the functional connectivity between the reconstructed regional time series. We used the weighted Minimum Norm Estimate (wMNE) to reconstruct the cortical sources and the phase locking value (PLV) method to compute the functional connectivity. This

This work was supported by the Rennes University Hospital (COREC Project named BrainGraph, 2015-17). The work has also received a French government support granted to the CominLabs excellence laboratory and managed by the National Research Agency in the "Investing for the Future" program under reference ANR-10-LABX-07-01. Authors would like to thank Dufor O. for his help in the data acquisition.

measure (range between 0 and 1) reflects interactions between oscillatory signals by quantifying their phase relationships. The PLVs were estimated at the gamma frequency band (30-45 Hz). The whole brain was parcellated into a set of 68 regions of interest identified by Desikan atlas [14]. The choice of wMNE/PLV was supported by two comparative analysis performed in [7][15].

*C. Analysis steps and practical issues*

In order to discriminate the brain connectivity networks related to each category, several steps were performed and summarized as following:

- **Step 1 – Brain network states:** to track the dynamics of the brain functional connectivity, we used a segmentation algorithm that allows decomposing the cognitive task into brain network states over time. This algorithm is based on the *k*-means clustering of the connectivity matrices, see [11] [12] for detailed description of the method. The averaged connectivity matrices over all subjects were first obtained. The application of the algorithm on the averaged connectivity matrix has led to six different functional connectivity clusters.

- **Step 2 - Functional connectivity matrix per picture:** The functional connectivity matrix for each picture was computed over available trials (in total 119 functional connectivity matrices were obtained). These matrices are segmented into six temporal windows as obtained in step 1.

- **Step 3 – Similarity between brain networks:** At each temporal window, the averaged connectivity matrix over time is computed for all pictures. The similarity score between all the connectivity networks for all pictures is computed to get 6 similarity matrices (119×119). The similarity score is computed using SimNet algorithm.

- **Step 4 – Modules in similarity matrices:** After calculating the similarity scores between the functional brain networks at each window, the Louvain modularity maximization method [16] was computed to every similarity matrix 100 times as the modularity index *Q* may vary from run to run, due to heuristic property of the algorithm. This produces a 119×119 matrix where values represent the probability of each two networks to be in the same module for all runs. We used Gephi [17] software to visualize the graph presentation of the modules, where the nodes represent the brain connectivity networks of pictures and the edges represent the similarity score between two brain connectivity networks of two different pictures.

## III. RESULTS

*A. Functional connectivity states and networks simialrity*

We apply the segmentation algorithm on the averaged connectivity matrices over all subjects, which led to 6 different clusters. The first cluster corresponds to the period ranging from t=0 (onset: presentation of the visual stimulus) to t=80 ms. The second cluster correspond to the period from 81ms to 195 ms. The third cluster from 196 ms to 276 ms, the fourth cluster correspond to the period from 277 ms to 384 ms, the fifth cluster correspond to the period from 285 ms to 460 ms and the sixth cluster from 461 ms to 620 ms. In order to measure the similarity between the meaningful and the meaningless pictures, the functional connectivity matrix for each picture was calculated to get 119 functional connectivity matrices (79 meaningful and 40 meaningless). SimNet was then applied between each pair of these networks, leading to a 119 x 119 similarity matrice. In figure 1, we show the results at the second brain network states (81- 195ms). The rows and the columns in this matrix represent the functional networks of the visual objects. Each voxel in this matrix represent a similarity score between each of two brain networks.

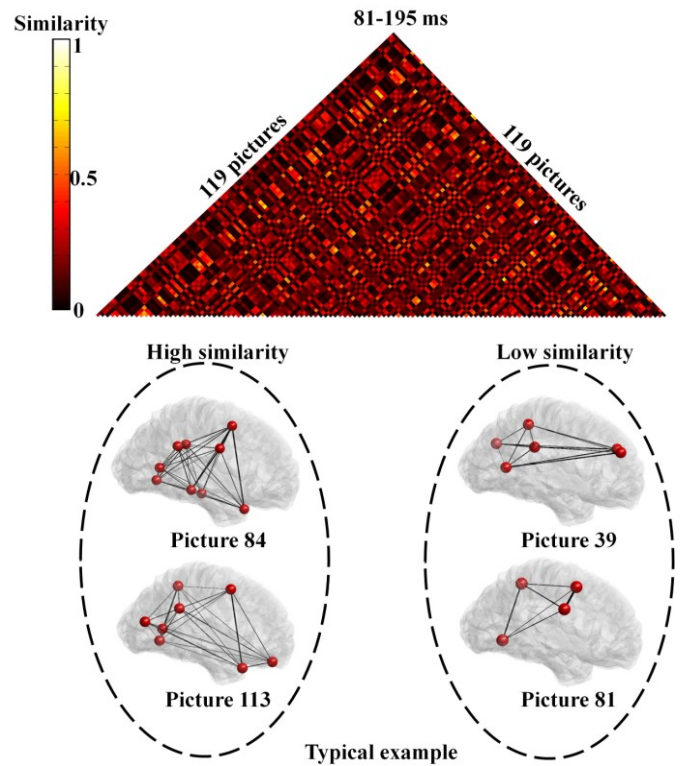

**Figure 1: UP:** Similarity matrix between brain networks identified from EEG during the second period (81- 195 ms) obtained from the segmentation algorithm, where the rows and columns represent the functional connectivity networks for all pictures (119 pictures). Bottom: two typical examples with 3D representation for two different similarity score (high similarity and low similarity) selected from the above similarity matrices.

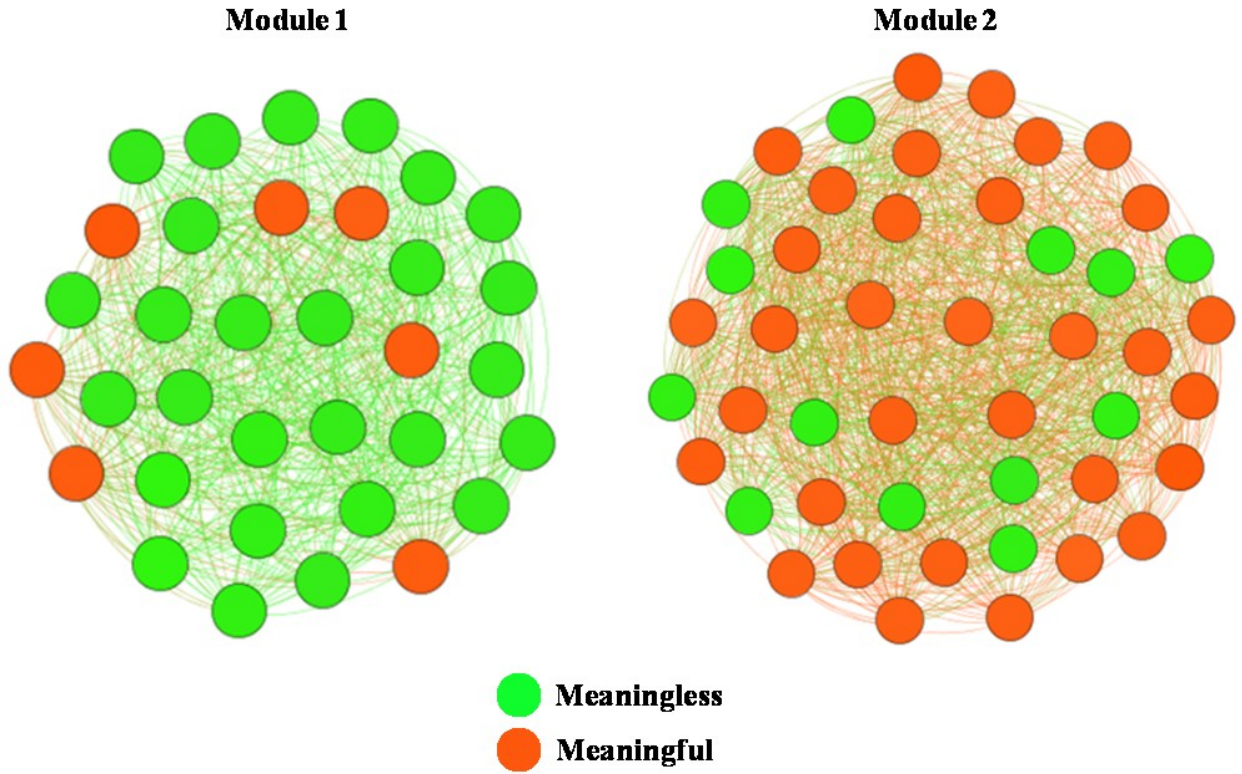

Figure 2: Results of application the modularity algorithm to the similarity matrix during the second period of the cognitive response.

In figure 1, we show also two typical examples of brain networks with high similarity values and low similarity values, where the node size represents the 'strength' of the network measure defined as the sum of weights of edges connected to this node. For instance, a high similarity was measured between the network obtained between two meaningless pictures (pictures 84 and 113). In contrast, a low similarity index was computed between the network obtained between one meaningful picture and another meaningless picture (pictures 39 and 81).

*B. Meaningful vs. meaningless*

Here, we apply the Louvain-algorithm on the similarity matrix where two main modules were detected. The graphical representations of these modules are presented in Figure 2 where the orange nodes represent meaningful pictures and the green nodes represent meaningless pictures. A global accuracy of 76% was obtained.

IV. DISCUSSION AND CONCLUSION

Understanding the information processing in the human brain and how the visual object categories are represented in the brain network is a big challenge. Several studies have found differences in the activated brain regions specified for object categories (animals, tools, faces…). Some of these studies have suggested that each category of objects is represented in a specific brain region [18]. Other studies showed that one brain region may represent many categories [19]. Huth et al. [20] used the brain activity recorded from subjects when watching natural movies, to study how the objects and the actions can be categorized in the brain. They found that the objects categories are regrouped in continuous semantic space across the brain areas.

In this paper, we present a new approach to differentiate between the visual object categories in the human brain based on brain network similarity. We combine both 'dense EEG source connectivity method' and network similarity algorithm 'SimNet' to investigate the categorization of the human brain networks. The combination was applied on visual task consist of naming pictures from two categories (meaningful and meaningless). To our knowledge, this study constitutes the first attempt to assess categorization in the human brain from a network-based approach between meaningful and meaningless pictures using dense-EEG recordings.

**Methodological considerations:** first a crucial parameter must be tuned in SimNet is the radius of the disk used to count the neighbors of a given node. An increase of the radius $R$ will automatically lead to an increase of the similarity index between the two compared graphs. In the application on the real data, $R$ was chosen as the average distance between all nodes ($R$=0.06). Second, it is worth mentioning here that the similarity scores between brain networks were calculated by taking the strongest 10% of edges in the connectivity matrices. Testing other threshold values or other threshold strategies is necessary to guarantee the robustness of the results. Third, we considered several values of the resolution parameter $\gamma$ (used to compute the modularity) and $\gamma$=1 was the good compromise between number for modules and nodes associated with each module (as reported also by many studies [21]).